\documentclass{aipproc}
\layoutstyle{6x9}
\usepackage{bm}
\usepackage{epsf}
\usepackage{rotating}
\begin{document}

\title{Heavy Quark Potential and Mass Spectra of Heavy Mesons}
\author{D. Ebert}{
  address={Institut f\"ur Physik, Humboldt--Universit\"at
zu Berlin, 10115 Berlin, Germany}}

\author{R. N. Faustov}{
address={Russian Academy of Sciences, Scientific Council for
Cybernetics, Moscow 117333, Russia}}

\author{V. O. Galkin}{
address={Russian Academy of Sciences, Scientific Council for
Cybernetics, Moscow 117333, Russia}}
\begin{abstract}
  The relativistic quark model is presented. The quark-antiquark
  potential for the Schr\"odinger-like equation is constructed with
  the account of retardation effects and one-loop radiative
  corrections. It consists of the one-gluon exchange part and the
  confining part which is the mixture of the Lorentz scalar and
  Lorentz vector contributions. The latter contains both the Dirac and
  Pauli terms. In the $v^2/c^2$ approximation the mass spectra of
  heavy quarkonia (charmonium and bottomonium) are calculated in good
  agreement with experiment. In the case of heavy-light mesons ($B$
  and $D$) the light quark is treated completely relativistically and
  only the expansion in the inverse heavy quark mass is used. The mass
  spectra of the ground and excited states of $D$, $D_s$, $B$, $B_s$
  mesons are calculated. They exhibit some features of the so-called
  ``level inversion''. The obtained results are generally in accord
  with experimental data. Still there exist some discrepancies between
  measurements of different collaborations   
\end{abstract}

\maketitle

\section{Introduction}
The heavy flavour studies lie on the frontiers of elementary particle
physics. The investigation of heavy meson mass spectra provides
substantial information about the nonperturbative content of quantum
chromodynamics (QCD) and fundamental parameters of the relativistic
quark model. Since it is impossible to give here any kind of
comprehensive review of the subject we present instead the
consideration based mainly on two papers \cite{efg,egf}. They are
extended to include recent experimental data and should be taken
as an illustration of theoretical approaches.

\section{ Relativistic quark model}
\label{sec:rqm}

In the quasipotential approach the meson is described by the wave
function of the bound quark-antiquark state, which satisfies the
quasipotential equation of the Schr\"odinger type in
the centre-of-mass frame:
\begin{equation}
\label{qpe}
{\left(\frac{b^2(M)}{2\mu_{R}}-\frac{{\bf
p}^2}{2\mu_{R}}\right)\Psi_{M}({\bf p})} =\int\frac{d^3 q}{(2\pi)^3}
 V({\bf p,q};M)\Psi_{M}({\bf q}),
\end{equation}
where $\mu_{R}$ is the relativistic reduced mass
$$
\mu_{R}=\frac{M^4-(m^2_q-m^2_Q)^2}{4M^3},
$$
and 
$b^2(M)$  denotes
the on-mass-shell relative momentum squared
$$
{b^2(M) }
=\frac{[M^2-(m_q+m_Q)^2][M^2-(m_q-m_Q)^2]}{4M^2}.
$$
The kernel
$V({\bf p,q};M)$ is the quasipotential operator of
the quark-antiquark interaction. It is constructed with the help of the
off-mass-shell scattering amplitude, projected onto the positive
energy states.

We have assumed that the effective
interaction is the sum of the usual one-gluon exchange term and the mixture
of vector and scalar linear confining potentials.
The quasipotential is then defined by
\begin{eqnarray}
\label{qpot}
V({\bf p,q};M)&=&\bar{u}_q(p)\bar{u}_Q(-p)\left\{\frac{4}{3}\alpha_sD_{ \mu\nu}({\bf
k})\gamma_q^{\mu}\gamma_Q^{\nu}
+V^V_{\rm conf}({\bf k})\Gamma_q^{\mu}
\Gamma_{Q;\mu}\right.\cr &&\left.\phantom{\frac{4}{3}}
+V^S_{\rm conf}({\bf k})({\bf p}, {\bf
q};M)\right\}u_q(q)u_Q(-q),
\end{eqnarray}
where $\alpha_s$ is the QCD coupling constant, $D_{\mu\nu}$ is the
gluon propagator in the Coulomb gauge
and ${\bf k=p-q}$; the Dirac spinor is given by
$$\label{spinor}
u^\lambda({p})=\sqrt{\frac{\epsilon(p)+m}{2\epsilon(p)}}
\left(\begin{array}{c}
1\\ \displaystyle\frac{\mathstrut\bm{\sigma}{\bf p}}{\mathstrut\epsilon(p)+m}
\end{array}\right)
\chi^\lambda
$$
with $\epsilon(p)=\sqrt{{\bf p}^2+m^2}$.
The effective long-range vector vertex has
the form
$$
\Gamma_{\mu}({\bf k})=\gamma_{\mu}+
\frac{i\kappa}{2m}\sigma_{\mu\nu}k^{\nu}, \qquad k^0=0,
$$
where $\kappa$ is  the 
nonperturbative anomalous chromomagnetic moment of quarks. Vector and
scalar confining potentials in the nonrelativistic limit reduce to
$$
V^V_{\rm conf}(r)=(1-\varepsilon)(Ar+B),\qquad
V^S_{\rm conf}(r) =\varepsilon (Ar+B),
$$
reproducing
$
V_{\rm conf}(r)=V^S_{\rm conf}(r)+
V^V_{\rm conf}(r)=Ar+B,
$
where $\varepsilon$ is the mixing coefficient.

\noindent The potential parameters are as follows:

$
A=0.18\ {\rm GeV}^2, \quad B=-0.30\ {\rm GeV}, \quad
\alpha_s(m_c)=0.32, \quad \alpha_s(m_b)=0.22$.

\noindent The constituent quark masses are

$m_b=4.88\ {\rm GeV}, \quad m_s=0.50\ {\rm GeV},\quad
m_c=1.55\ {\rm GeV}, \quad m_{u,d}=0.33\ {\rm GeV}$.

\noindent The value of the mixing parameter $\varepsilon=-1$
is fixed by matching the  heavy quark expansion
and from the consideration of charmonium radiative decays ($J/\psi\to
\eta_c\gamma$). 
The anomalous chromomagnetic quark moment $\kappa=-1$ is determined
from the heavy quark expansion
and from fitting the fine splitting of heavy quarkonia ${}^3P_J$ states.
The long range chromomagnetic contribution to the 
potential is proportional to $(1+\kappa)$ and vanishes for
$\kappa=-1$ in accord with the flux tube model.

\section{Heavy quark-antiquark potential}
\label{sec:hqp}

With the account of retardation 
effects and one loop radiative corrections the spin-independent
potential reads as \cite{efg} 
\begin{eqnarray}
\label{sipot}
V_{\rm SI}(r)
&=&-\frac43\frac{\bar \alpha_V(\mu^2)}{r} +Ar+B -\frac43\frac{\beta_0
\alpha_s^2(\mu^2)}{2\pi}\frac{\ln(\mu r)}{r} \cr
&& +\frac18\left(\frac{1}{m_a^2}+\frac{1}{m_b^2}\right) \Delta\Biggl[ -\frac43\frac{\bar 
\alpha_V(\mu^2)}{r} -\frac43\frac{\beta_0\alpha_s^2(\mu^2)}{2\pi}\frac{\ln(\mu r)}{r}\cr
&& +(1-\varepsilon)(1+2\kappa)Ar\Biggr]
+\frac{1}{2m_am_b}\Biggl(\left\{-\frac43\frac{\bar\alpha_V}{r}
\left[{\bf p}^2+\frac{({\bf p\cdot r})^2}{r^2}\right]\right\}_W\cr
&&-\frac43\frac{\beta_0\alpha_s^2(\mu^2)}{2\pi}\left\{{\bf p}^2\frac{\ln(\mu r)}{r}
+\frac{({\bf p\cdot 
r})^2}{r^2}\left(\frac{\ln(\mu r)}{r}-\frac1r\right)\right\}_W\Biggr)\cr
&& +\left[\frac{1-\varepsilon}{2m_am_b}-\frac{\varepsilon}{4}
\left(\frac{1}{m_a^2}+\frac{1}{m_b^2}\right)\right]\left\{Ar\left[{\bf p}^2
-\frac{({\bf p\cdot r})^2}{r^2}\right]\right\}_W\cr
&&-\frac{\varepsilon\lambda_S}{2}\left[\frac12
\left(\frac{1}{m_a^2}+\frac{1}{m_b^2}\right)+\frac{1}{m_am_b}\right]
\left\{Ar\left[{\bf p}^2+\frac{({\bf p\cdot r})^2}{r^2}\right]\right\}_W\cr
&&+\left[\frac{1}{4}\left(\frac{1}{m_a^2}+\frac{1}{m_b^2}\right)+
\frac{1}{m_am_b}\right]B{\bf p}^2,
\end{eqnarray}
where
\begin{eqnarray*}
\bar\alpha_V(\mu^2)&=&\alpha_s(\mu^2)\left[1+\left(\frac{a_1}{4}
+\frac{\gamma_E\beta_0}{2}\right)\frac{\alpha_s(\mu^2)}{\pi}\right],\\
\alpha_s(\mu^2)&=&\frac{4\pi}{\beta_0\ln(\mu^2/\Lambda^2)},\qquad
a_1=\frac{31}{3}-\frac{10}{9}n_f,\qquad \beta_0=11-\frac23n_f.\nonumber
\end{eqnarray*}
Here $n_f$ is a number of flavours, $\mu$ is a renormalization scale
and the subscript $W$ means the Weyl ordering of operators.

The spin-dependent part of the quark-antiquark potential for equal
quark masses ($m_a=m_b=m$) is given by \cite{efg}
\begin{eqnarray}
\label{vsd}
V_{\rm SD}(r)&
=& a\ {\bf L}\cdot{\bf S}+b\left[\frac{3}{r^2}({\bf S}_a\cdot {\bf r})
({\bf S}_b\cdot {\bf r})-({\bf S}_a\cdot {\bf S}_b)\right] +c\ {\bf S}_a\cdot {\bf S}_b, \\
\label{a}
a&=& \frac{1}{2m^2}\Biggl\{\frac{4\alpha_s(\mu^2)}{r^3}\Biggl(1+
\frac{\alpha_s(\mu^2)}{\pi}\Biggl[\frac{1}{18}n_f-\frac{1}{36}+\gamma_E\left(
\frac{\beta_0}{2}-2\right)+\frac{\beta_0}{2}\ln\frac{\mu}{m}\cr
&&+\left(\frac{\beta_0}{2}-2\right)\ln(mr)\Biggr]\Biggr)
-\frac{A}{r}+4(1+\kappa)(1-\varepsilon)\frac{A}{r}\Biggr\},\cr
\label{b}
b&=& \frac{1}{3m^2}\Biggl\{\frac{4\alpha_s(\mu^2)}{r^3}\Biggl(1+
\frac{\alpha_s(\mu^2)}{\pi}\Biggl[\frac{1}{6}n_f+\frac{25}{12}+\gamma_E\left(
\frac{\beta_0}{2}-3\right)+\frac{\beta_0}{2}\ln\frac{\mu}{m}\cr
&&+\left(\frac{\beta_0}{2}-3\right)\ln(mr)\Biggr]\Biggr)
+(1+\kappa)^2(1-\varepsilon)\frac{A}{r}\Biggr\},\cr
\label{c}
c&=& \frac{4}{3m^2}\Biggl\{\frac{8\pi\alpha_s(\mu^2)}{3}\Biggl(\left[1+
\frac{\alpha_s(\mu^2)}{\pi}\left(\frac{23}{12}-\frac{5}{18}n_f-\frac34\ln2\right)
\right]\delta^3(r)\cr
&&+\frac{\alpha_s(\mu^2)}{\pi}\left[-
\frac{\beta_0}{8\pi}\nabla^2\left(\frac{\ln({\mu}/{m})}{r}\right)
+\frac{1}{\pi}\left(\frac{1}{12}n_f-\frac{1}{16}\right)\nabla^2\left(
\frac{\ln(mr)+\gamma_E}{r}\right)\right]\Biggr)\cr
&&+(1+\kappa)^2(1-\varepsilon)\frac{A}{r}\Biggr\},\nonumber
\end{eqnarray}
where ${\bf L}$ is the orbital momentum and ${\bf S}_{a,b}$, ${\bf
  S}={\bf S}_a+ {\bf S}_b$ are the spin momenta. The total angular
  momentum is $\bf J=\bf L+\bf S$.

\section{Heavy quarkonium mass spectra}
\label{sec:hqms}

The heavy quarkonium is similar to the positronium atom and its levels
are usually specified by the notation $n^{(2S+1)}L_J$, where $n$ is the
radial quantum number. The results of calculations of heavy quarkonium
mass spectra on the basis of Eqs.~(\ref{qpe}), (\ref{sipot}),
(\ref{vsd}) are presented in Tables 1 and 2 \cite{efg}.

Recently the contribution of the finite $c$-quark mass to the
bottomonium mass spectrum in one loop was considered in our model
\cite{efg3}. The correction to the $b\bar b$ static poten-

\begin{table}[bh]
\caption{Charmonium Mass Spectrum (GeV). }
\bigskip
\label{charm}
\begin{tabular}{ccclccc}
\hline
State 
&Particle &  Theory  & PDG(2000)&
 BES &CLEO& E835 \\
($n^{(2S+1)}L_J$)&&\cite{efg} &\ \ \ \  \ \cite{pdg}&&\\
\hline
$1^1S_0$& $\eta_c$ & 2.979 & 2.9798(18) &2.9763&2.9804 &\\
$1^3S_1$& $J/\Psi$ & 3.096 & 3.09687(4) &&&\\
&&&&&&\\
$1^3P_0$& $\chi_{c0}$ & 3.424 & 3.4150(8) & 3.4141&&3.4154\\ 
$1^3P_1$& $\chi_{c1}$ & 3.510 & 3.51051(12) &&&\\
$1^3P_2$& $\chi_{c2}$ & 3.556 & 3.55618(13) &&&\\
&&&&&&\\
$2^1S_0$& $\eta_c'$ & 3.583 & 3.594(5) &&&\\
$2^3S_1$& $\Psi'$     & 3.686 & 3.68596(9) &&&\\ 
&&&&&&\\
$1^3D_1$&  & 3.798 & $3.7699(25)^*$ &&&\\
$1^3D_2$&  & 3.813 & &&&\\
$1^3D_3$&  & 3.815 & &&&\\
&&&&&&\\
$2^3P_0$& $\chi'_{c0}$ & 3.854 & &&&\\
$2^3P_1$& $\chi'_{c1}$ & 3.929 & &&&\\
$2^3P_2$& $\chi'_{c2}$ & 3.972 & &&&\\
&&&&&&\\
$3^1S_0$& $\eta_c''$ & 3.991 & &&&\\
$3^3S_1$& $\Psi''$     & 4.088 & $4.040(10)$\tablenote{Mixture of $S$
  and $D$ states}  &&&\\
&&&&&&\\
$2^3D_1$&  & 4.194 & $4.159(20)^*$ &&&\\
$2^3D_2$&  & 4.215 & &&&\\
$2^3D_3$&  & 4.223 & &&&\\ 
\hline
\end{tabular}
\end{table} 

\begin{table}[t]
\caption{Bottomonium Mass Spectrum (GeV). }
\bigskip
\label{bottom}
\begin{tabular}{ccllcc}
\hline
State & Particle &  Theory & PDG(2000)
& CLEO& MD1\\
($n^{(2S+1)}L_J$)&&\phantom{aa}\cite{efg}&\phantom{aaa}\cite{pdg}&&\\
\hline
$1^1S_0$& $\eta_b$ & 9.400 &  &&\\
$1^3S_1$& $\Upsilon$ & 9.460 & 9.46030(26) &&9.46051\bigskip\\
$1^3P_0$& $\chi_{b0}$ & 9.864 & 9.8599(10) & 9.8600&\\
$1^3P_1$& $\chi_{b1}$ & 9.892 & 9.8927(6) & 9.8937&\\
$1^3P_2$& $\chi_{b2}$ & 9.912 & 9.9126(5) &  9.9119&\bigskip\\
$2^1S_0$& $\eta_b'$ & 9.990 & &&\\
$2^3S_1$& $\Upsilon'$  & 10.020 & 10.02326(31) &&10.0235\\ 
$1^3D_1$&  & 10.151 &  & &\\
$1^3D_2$&  & 10.157 &  &&\\
$1^3D_3$&  & 10.160 & &&\bigskip\\
$2^3P_0$& $\chi'_{b0}$ & 10.232 & 10.2321(6) & &\\
$2^3P_1$& $\chi'_{b1}$ & 10.253 & 10.2552(5) & &\\
$2^3P_2$& $\chi'_{b2}$ & 10.267 & 10.2685(4) & &\bigskip\\
$3^1S_0$& $\eta_b''$ & 10.328 & & &\\
$3^3S_1$& $\Upsilon''$ & 10.355 & 10.3552(5) &&\bigskip\\
$2^3D_1$&  & 10.441 &   & &\\
$2^3D_2$&  & 10.446 &  &&\\
$2^3D_3$&  & 10.450 &  &&\bigskip\\ 
$3^3P_0$& $\chi''_{b0}$ & 10.498 &  & &\\
$3^3P_1$& $\chi''_{b1}$ & 10.516 &  & &\\
$3^3P_2$& $\chi''_{b2}$ & 10.529 &  & &\bigskip\\
$4^1S_0$& $\eta_b'''$ & 10.578 & & &\\
$4^3S_1$& $\Upsilon'''$ & 10.604 & 10.5800(35) &&\\
\hline
\end{tabular}
\end{table}

\noindent tial due to
$m_c\not=0$ is approximately given by \cite{m} \ ($a_0=5.2$,\quad
$\gamma_E=0.5772\dots$) 
$$ \delta V(r)\cong -\frac49\frac{\alpha_s^2}{\pi
  r}\left[\ln(\sqrt{a_0}m_c r)+\gamma_E+{\rm E}_1(\sqrt{a_0}m_c
  r)\right],  \qquad {\rm E}_1(x)=\int_x^\infty \frac{{\rm d}t}te^{-t}.$$
Averaging over solutions of Eq.~(\ref{qpe}) with the Cornell
potential yields the following bottomonium mass shifts:
\begin{tabular}{crrrrrr}
State& $1S$ & $1P$ & $2S$ & $1D$ & $2P$ & $3S$\\
\hline
$\langle\delta V\rangle$, MeV & $-12$ & $-9.3$ & $-8.7$ & $-7.6$ &
$-7.5$ & $-7.2$
\end{tabular}\ .
These shifts are within the estimates of theoretical uncertainties
($\sim 10$ MeV) and partially could be adsorbed either in the value of
the constituent $b$ quark mass or in the constant term $B$ of the
confining potential.

\section{Mass Spectra of $\bm{B}$ and $\bm{D}$ mesons}
\label{sec:msm}

The $B$ and $D$ mesons are alike the hydrogen atom with the heavy
quark $Q=b,c$ near the centre-of-mass and the light quark $q=u,d,s$
orbiting around it. Thus it is appropriate to make the expansion in
the inverse heavy quark mass. In the limit of infinitely heavy quark
($m_Q\to\infty$) its mass and spin decouple and as a result heavy
quark symmetry arises. In this limit the meson angular momentum is a
sum of the orbital momentum ${\bf L}$ and the light quark spin ${\bf
  S}_q$ i.e. ${\bf j}={\bf L}+{\bf S}_q$. The $1/m_Q$ correction to
the $Q\bar q$ potential depends on the heavy quark spin ${\bf S}_Q$
and leads to the spin-spin interaction. The total angular momentum is
${\bf J}={\bf j}+{\bf S}_q$. Thus, for $S$-wave mesons there is a
doublet of $j=1/2$ states with $J^P=0^-,1^-$ ($P$ is the meson parity)
which are degenerate in the limit $m_Q\to\infty$. For $P$-wave mesons
there are similarly two doublets of initially degenerate states with
$j=1/2$ ($J^P=0^+,1^+$) and $j=3/2$ ($J^P=1^+,2^+$). The $j=1/2$
levels are expected to be broad because they decay in an $S$-wave,
while the $j=3/2$ levels should be narrow since they decay in a
$D$-wave.

The heavy-light quark-antiquark potential
in configuration space in the limit $m_Q\to\infty$ reads as \cite{egf}
($V_{\rm Coul}(r)=-(4/3)\alpha_s/r$,\quad $\alpha_s=0.5$ for $B, D$;\ \
$\alpha_s=0.45$ for $B_s, D_s$) 
\begin{eqnarray}
\label{vinf}
V_{m_Q\to \infty}(r)
&=& \frac{E_q+m_q}{2E_q}\Bigg[V_{\rm Coul}(r)
+V_{\rm conf}(r) + \frac{1}{(E_q+m_q)^2}\Bigg\{{\bf p}[\tilde 
V_{\rm Coul}(r)\cr
& & +V_{\rm conf}^V(r)-V_{\rm conf}^S(r)]{\bf p}
-\frac{E_q+m_q}{2m_q}\Delta V_{\rm
conf}^V(r)[1-(1+\kappa)] \\
& &+\frac{2}{r}\left(\tilde V_{\rm Coul}'(r)-V_{\rm conf}'^S(r) -
V_{\rm conf}'^V(r)\left[\frac{E_q}{m_q}
-2(1+\kappa)\frac{E_q+m_q}{2m_q}\right]\right)
{\bf L}{\bf S}_q\Bigg\}\Bigg].\nonumber
\end{eqnarray}
The $1/m_Q$ correction to the potential (\ref{vinf}) is given by \cite{egf}
\begin{eqnarray}
\label{vcor}
\delta V_{1/m_Q}(r)
&\!\!\!\!=\!\!\!\!&\frac{1}{E_qm_Q}\Bigg\{{\bf p}\left[V_{\rm
Coul}(r)+V^V_{\rm conf}(r)\right]{\bf p}
+V'_{\rm Coul}(r)\frac{{\bf L}^2}
{2r}\cr
&&-\frac{1}{4}\Delta V^V_{\rm conf}(r)+\left[\frac{1}{r}V'_{\rm 
Coul}(r)+\frac{(1+\kappa)}{r}V'^V_{\rm conf}(r)\right]{\bf LS}\cr
& & +\frac13\biggl(\frac{1}{r}V'_{\rm Coul}(r)-V''_{\rm Coul}(r)
+(1+\kappa)^2\left[\frac{1}{r}V'^V_{\rm conf}(r)-V''^V_{\rm conf}(r)
\right]\biggr)\\
&&\times
\left[-{\bf S}_q{\bf S}_Q+\frac{3}{r^2}({\bf S}_q{\bf
r})({\bf S}_Q{\bf r})\right]
+\frac23\left[\Delta V_{\rm Coul}(r)+(1+\kappa)^2\Delta V^V_{\rm 
conf}(r)\right]{\bf S}_Q{\bf S}_q\Bigg\}. \nonumber
\end{eqnarray}
Here the prime denotes differentiation with respect to $r$, ${\bf L}$ is
the orbital momentum, 
${\bf S}_q$ and ${\bf S}_Q$ are the spin operators of the light and heavy 
quarks, ${\bf S}={\bf S}_q+{\bf S}_Q$ is the total spin. 

First Eq.~(\ref{qpe}) is solved numerically with the complete
potential (\ref{vinf}) and then the corrections (\ref{vcor}) is
treated perturbatively. The results of the calculation of heavy-light
meson mass spectra are presented in Tables~3-6 and in Figs.~1-4 \cite{egf}.

\begin{table}
\caption{ Mass Spectrum of $D$ Mesons (GeV).}

\begin{tabular}{cccllc}
\hline
 State&Particle& Theory \cite{egf} & PDG(2000) &\phantom{aaaaa}CLEO& DELPHI\\
\hline
$1S_0$&$D$ & 1.875 &  1.8693(5)& &\\
$1S_1$&$D^*$ & 2.009 & 2.0100(5)& &\bigskip\\
$1P_2$&$D_2^*$ & 2.459 & 2.459(4)& &\\
$1P_1$&$D_1$ &2.414 & 2.4222(18)&2.425$\pm0.002\pm0.002$& \\
$1P_1$&$D_1$ & 2.501 &  &{ 2.461}$^{+0.041}_{-0.034}{\pm0.01\pm0.032}$&     \\
$1P_0$&$D_0^*$ & 2.438 & &      &\bigskip\\
$2S_0$&$D'$ & 2.579 &       &      & \\
$2S_1$&$D^{*'}$ & 2.629  &&     & 2.637(9) ? \\
\hline
\end{tabular}
\end{table}

\begin{table}
\caption{ Mass Spectrum of $D_s$ Mesons (GeV).}
\begin{tabular}{ccccc }
\hline
State&\phantom{aaa} Particle\phantom{aaa}&\phantom{aaa}
Theory \cite{egf} &\phantom{aaa} PDG(2000)\phantom{aaa}&
\phantom{aaa} FOCUS\phantom{aaa}  \\
\hline
$1S_0$&$D_s$ & 1.981 & 1.9686(6)&\\
$1S_1$&$D_s^*$ & 2.111 & 2.1124(7)&\bigskip\\
$1P_2$&$D_{s2}^*$ & 2.560 & 2.5735(17)& 2.5673(13) \\
$1P_1$&$D_{s1}$ & 2.515 & 2.53535(60)& 2.5351(6) \\
$1P_1$&$D_{s1}$ & 2.569 &    &   \\
$1P_0$&$D_{s0}^*$ & 2.508 &   &   \bigskip \\
$2S_0$&$D_s'$ & 2.670 &     &  \\
$2S_1$&$D_s^{*'}$ & 2.716 &  &     \\
\hline
\end{tabular}
\end{table}

\begin{table}[ht]
\caption{ Mass Spectrum of $B$ Mesons (GeV).}\bigskip

\begin{tabular}{ccccccccc}
\hline
State& Part. & Theor.\cite{egf} & PDG(2000)& OPAL& L3& DELPHI& CDF& ALEPH\\
\hline
$1S_0$&$B$ & 5.285 & 5.2790(5)& & & & &\\
$1S_1$&$B^*$ & 5.324 & 5.3250(6)& & & &\bigskip \\
$1P_2$&$B^*_2$ & 5.733 & & &5.768(8)? &5.732(21)& &5.739(13) \\
$1P_1$&$B_1$ & 5.719 & & 5.738(9)& & & 5.71(2)&        \\
$1P_1$&$B_1$ & 5.757 & & & 5.670(16)?& & &     \\
$1P_0$&$B_0^*$ & 5.738 & &5.839(14)& & & &       \bigskip  \\
$2S_0$&$B'$ & 5.883 & & & & & &       \\
$2S_1$&$B^{*'}$ & 5.898 & & & & 5.90(2) ?& & \\
\hline
\end{tabular}
\end{table}
 
\begin{table}
\caption{ Mass Spectrum of $B_s$ Mesons (GeV).}

\begin{tabular}{ccccc}
\hline
 State&\phantom{aa} Particle\phantom{aa} &\phantom{aa}
 Theory \cite{egf}  &\phantom{aa}  PDG(2000)\phantom{aa}&\phantom{aa} OPAL\phantom{aaa}\\
\hline
$1S_0$&$B_s$ & 5.375 & 5.3696(24)&\\
$1S_1$&$B_s^*$ & 5.412 & 5.4166(35)& \bigskip \\
$1P_2$&$B_{s2}^*$ & 5.844 & & 5.853(15)  \\
$1P_1$&$B_{s1}$ & 5.831 &  &       \\
$1P_1$&$B_{s1}$ & 5.859 & &         \\
$1P_0$&$B_{s0}^*$ & 5.841 & &         \bigskip \\
$2S_0$&$B_s'$ & 5.971 & &          \\
$2S_1$&$B_s^{*'}$ & 5.984 & &          \\
\hline
\end{tabular}
\end{table}

\begin{figure}[hbt]

\centerline{\begin{turn}{-90}\epsfxsize=8.5cm 
\epsfbox{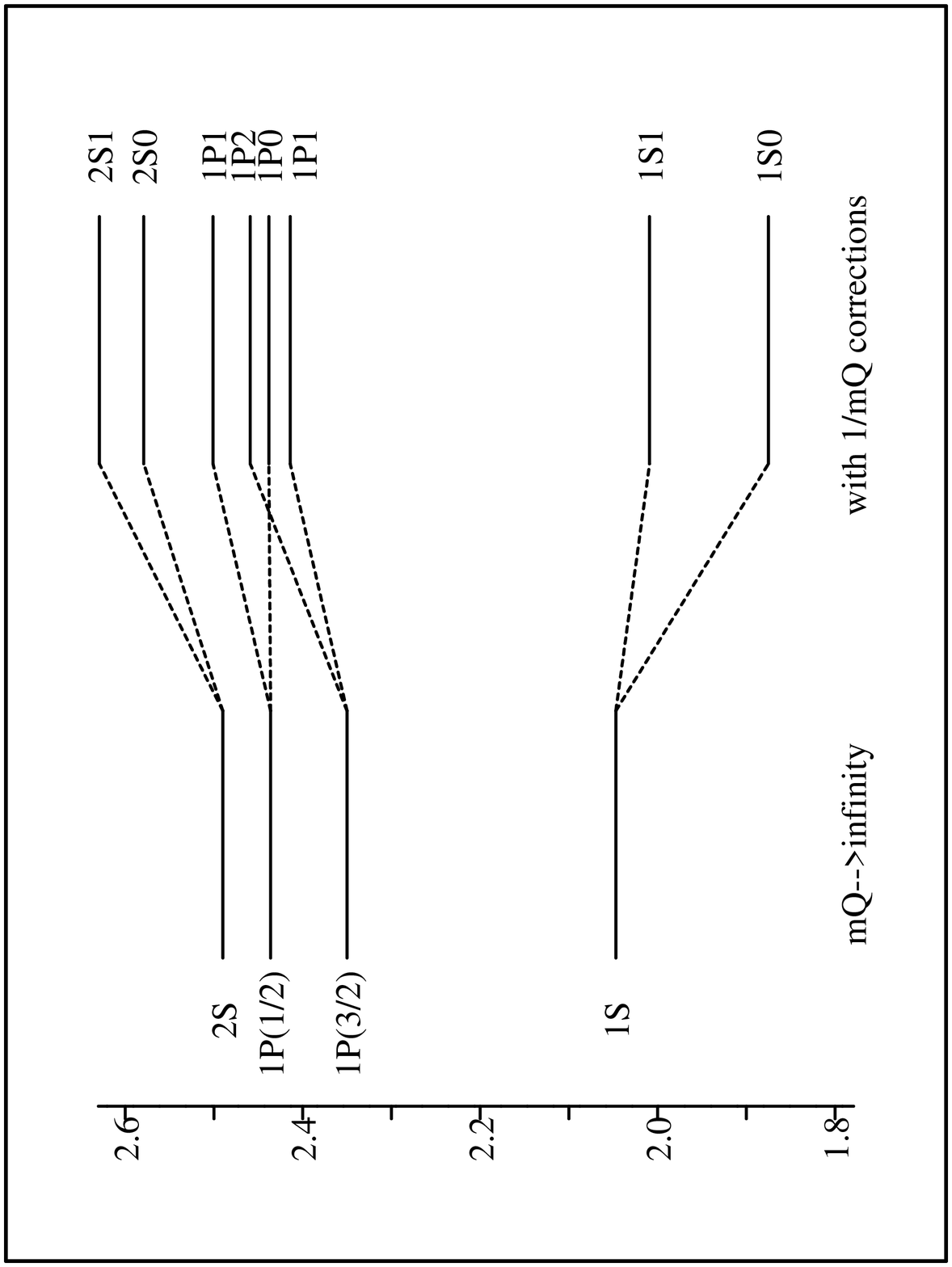}\end{turn}}

\caption{The ordering pattern of $D$ meson states. The mass scale is in GeV.}
\end{figure}

\begin{figure}[hbt]

\centerline{\begin{turn}{-90}\epsfxsize=8.5cm 
\epsfbox{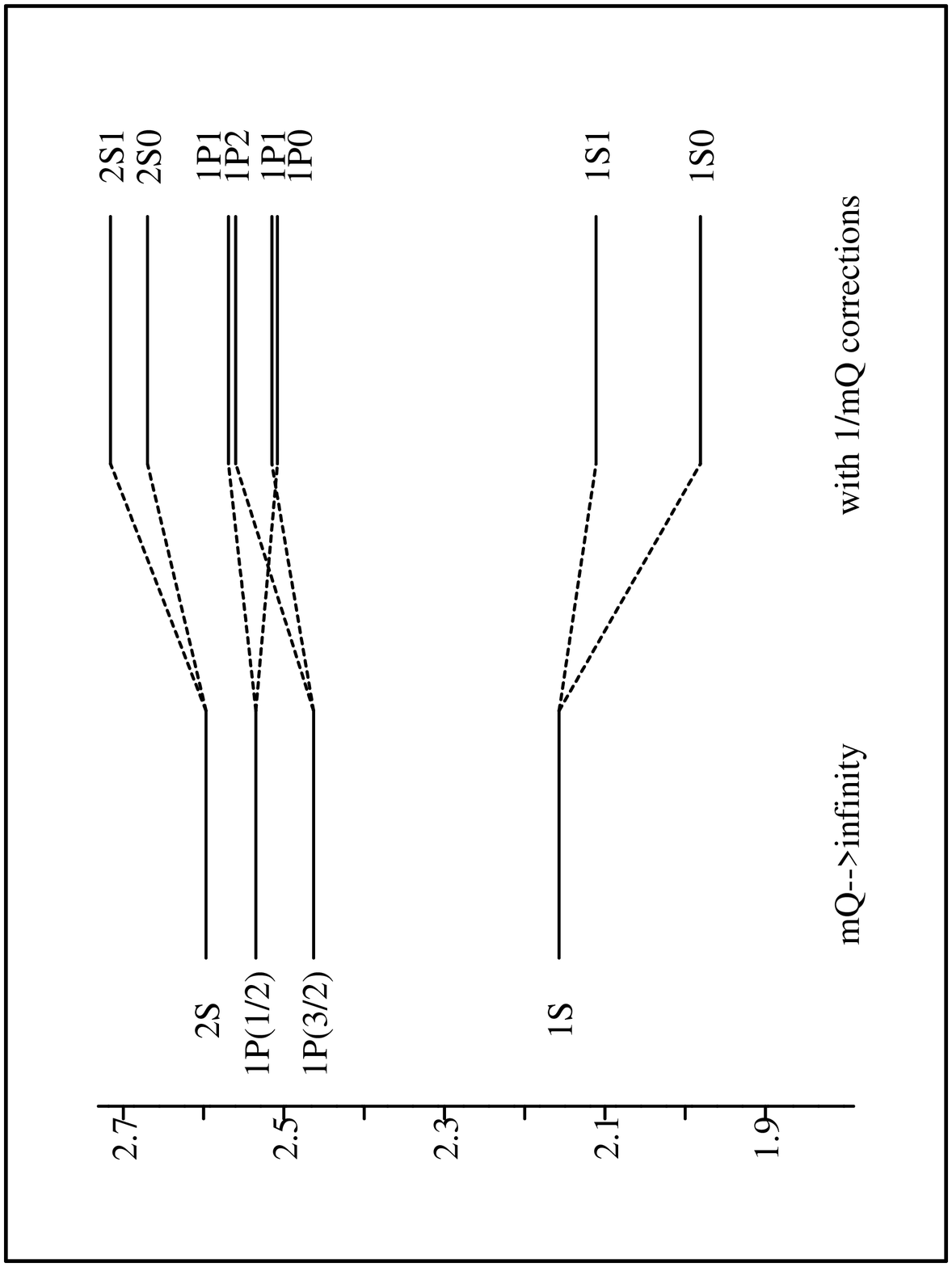}\end{turn}}

\caption{The ordering pattern of $D_s$ meson states. The mass scale is in GeV.}
\label{ds}
\end{figure}

\begin{figure}[hbt]

\centerline{\begin{turn}{-90}\epsfxsize=8.5cm 
\epsfbox{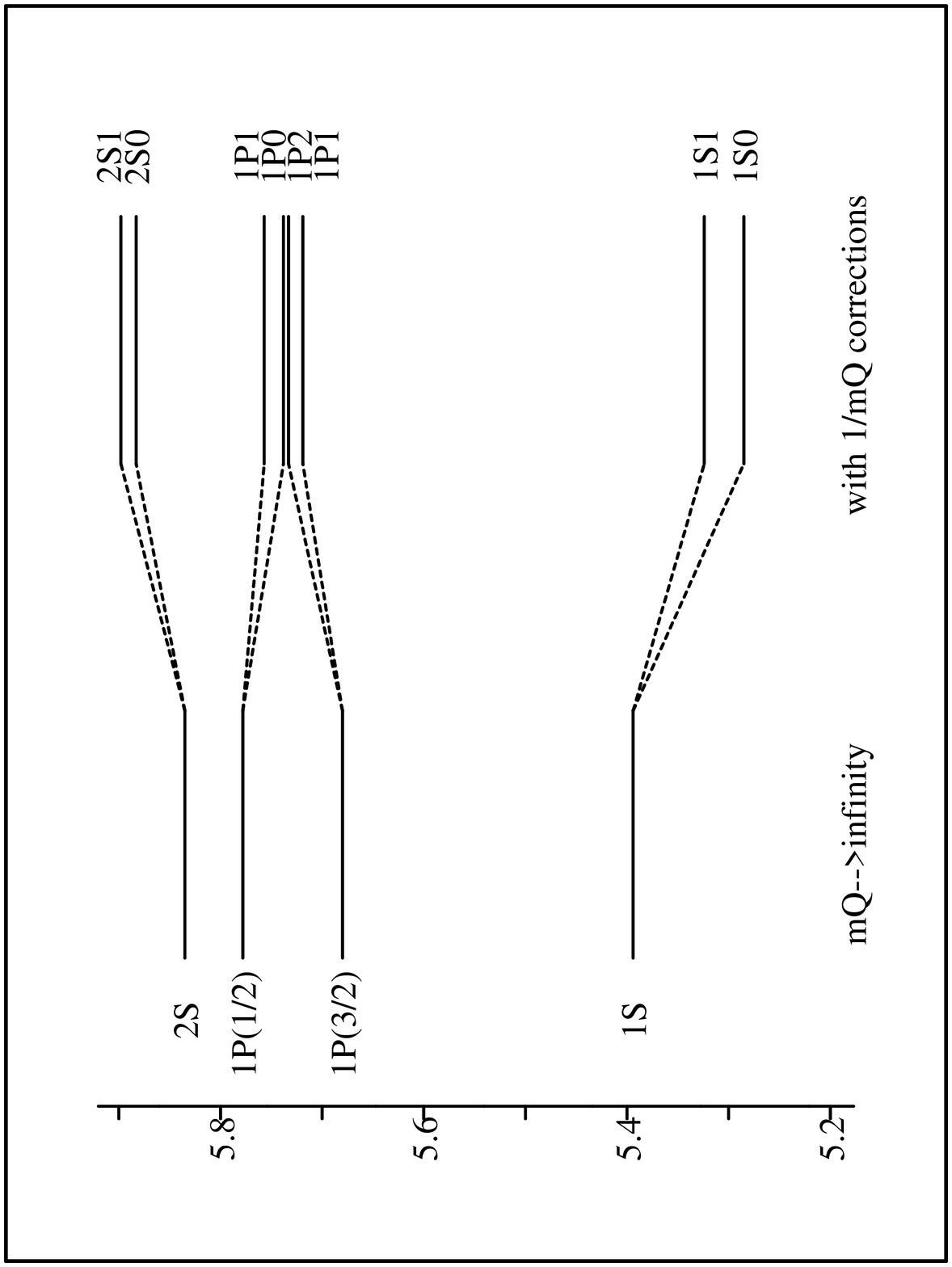}\end{turn}}

\caption{The ordering pattern of $B$ meson states. The mass scale is in GeV.}
\end{figure}

\begin{figure}[hbt]

\centerline{\begin{turn}{-90}\epsfxsize=8.5cm 
\epsfbox{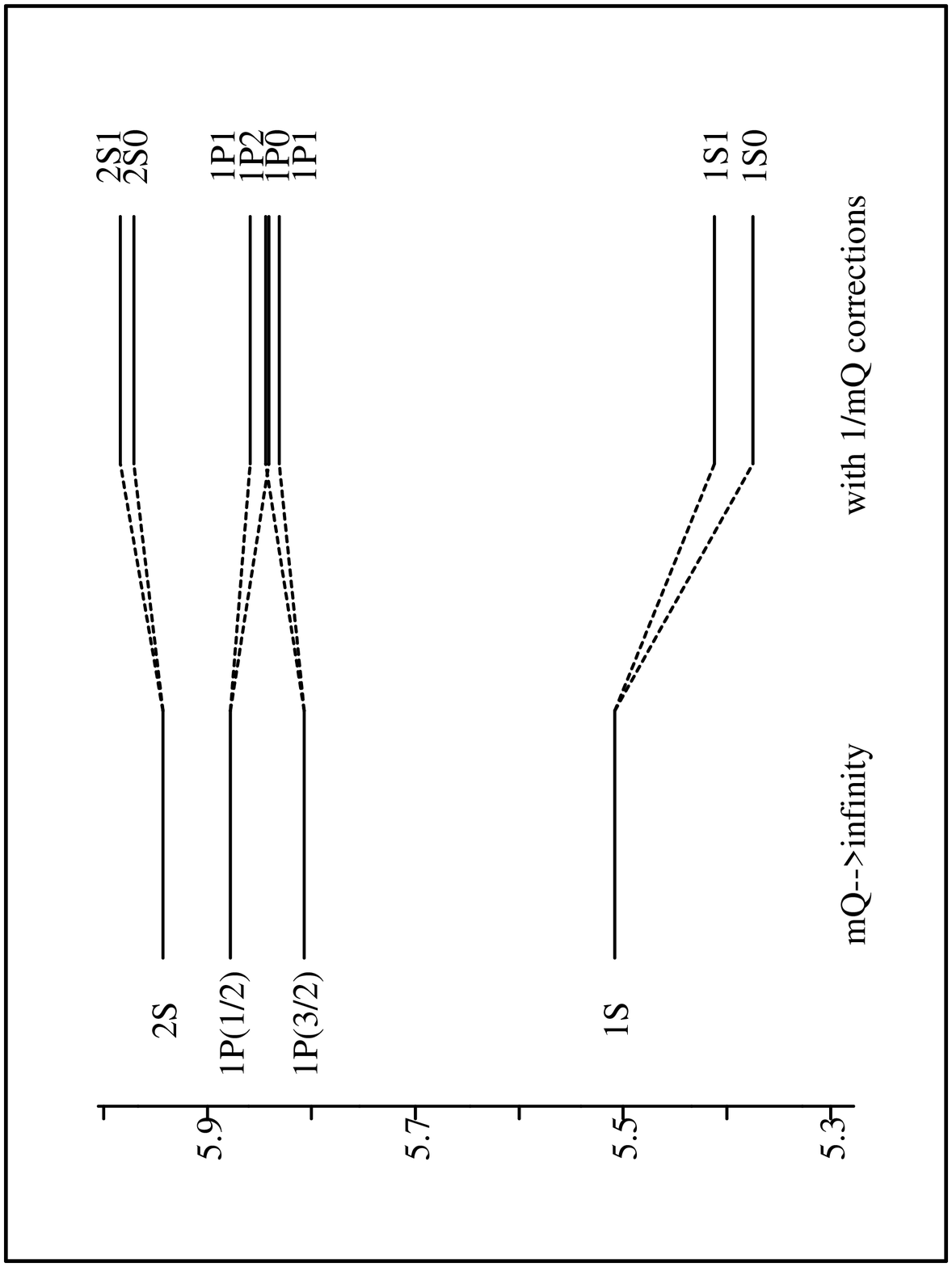}\end{turn}}

\caption{The ordering pattern of $B_s$ meson states. The mass scale is 
in GeV.}
\end{figure}

From Figs.~1-4 it follows that in the limit $m_Q\to\infty$ the
$P$-wave doublet with $j=1/2$ lies higher than the one with $j=3/2$
(abnormal level ordering), i.e. these doublets are inverted.
For finite $m_Q$ the hyperfine splitting of the doublet states makes
this picture much more complicated (some of the levels from different
doublets overlap). Only for $B$ mesons the pure inversion is preserved.

In the limit $m_Q\to\infty$ heavy quark symmetry predicts simple
relations between the spin-averaged masses of $B$ and $D$ states
\begin{equation}
  \label{eq:sr}
  \bar M_{B_1}-\bar M_{D_1}=\bar M_{B_{s1}}-\bar M_{D_{s1}}=\bar
  M_{B_s}-\bar M_{D_s}=\bar M_{B}-\bar M_{D}=m_b-m_c=3.33\ {\rm GeV}, 
\end{equation}
where $\bar M_{B_1}=(3M_{B_1}+5M_{B_2})/8$, $\bar
M_B=(M_B+3M_{B^*})/4$, etc. From Tables~3-6 the following values (in
GeV)  for the calculated mass differences can be obtained
\begin{center}
\begin{tabular}{cccc}
$\bar M_{B_1}-\bar M_{D_1}$&$\bar M_{B_{s1}}-\bar M_{D_{s1}}$ &$\bar
  M_{B_s}-\bar M_{D_s}$& $\bar M_{B}-\bar M_{D}$\cr
3.29&3.30&3.34&3.33
\end{tabular}  
\end{center}
Thus relations (\ref{eq:sr}) are satisfied with good accuracy.

The hyperfine mass splittings of the initially degenerate $P$ states
$$\Delta M_B\equiv M_{B_2}- M_{B_1},\ M_{B_1}- M_{B_0}; \qquad
\Delta M_D\equiv M_{D_2}- M_{D_1},\ M_{D_1}- M_{D_0}$$
should scale with heavy quark masses: $\Delta M_B=(m_c/m_b)\Delta M_D$
and the same for $B_s$ and $D_s$ mesons. Our model predictions for
these splittings are displayed in Table 7 \cite{egf}.
\begin{table}
\caption{ Hyperfine Splittings of $P$ Levels (MeV).}
\label{splt}
\begin{tabular}{cccc|ccc}
\hline
 States & $\Delta M_D$ & \phantom{aa}$\frac{m_c}{m_b}\Delta M_D$ & $\Delta M_B$ &
$\Delta M_{D_s}$ &\phantom{a} $\frac{m_c}{m_b}\Delta M_{D_s}$ & $\Delta M_{B_s}$\\
\hline
$1P_2-1P_1$ & 45 & 14 & 14 & 45 & 14 & 13\\
$1P_1-1P_0$ & 63 & 20 & 19 & 61 & 19 & 18\\
\hline
\end{tabular}
\end{table}

\section{Conclusions}
\label{sec:c}

So one may conclude that for heavy quarkonium mass spectra most of
theoretical calculations are generally in good agreement with
high-precision experimental data. This makes possible an accurate
determination of fundamental parameters governing the dynamics of the
heavy $Q\bar Q$ interaction. On the other hand, for heavy-light meson
mass spectra the situation is much more indefinite and unclear in both
theory and experiment. Not all models predict the $P$ level
inversion. The broad $j=1/2$ levels are very poorly studied due to
difficulties in their observation. Much further work is required to
remove discrepancies between existing experimental data. 

This research was supported in part by the {\it Deutsche
Forschungsgemeinschaft} under  Contract Eb 139/2-1, {\it
 Russian Foundation for Fundamental Research} under Grant No.\
00-02-17768 and {\it Russian Ministry of Education} under Grant
No. E00-3.3-45.

\end{document}